\documentclass[journal,onecolumn]{IEEEtran}
\ifCLASSINFOpdf

\else

\fi

\usepackage[T1]{fontenc}
\usepackage{amsmath,amssymb,amsfonts}
\usepackage{graphicx}
\usepackage{textcomp}
\usepackage{xcolor}
\usepackage{multirow}
\usepackage{booktabs}
\usepackage{array}
\usepackage{url}
\usepackage{setspace}
\usepackage{caption}
\usepackage{tabularx}
\newcolumntype{Y}{>{\raggedright\arraybackslash}X}
\usepackage{siunitx}
\usepackage{lineno}
\usepackage{mathptmx}            
\usepackage[super,comma,sort&compress]{natbib}  
\usepackage[hidelinks]{hyperref}
\begin{document}

\title{Beyond Silica Assumptions: Optical Network Design in the	Hollow-Core Era}

\author{Md Ghulam Saber and Zhiping Jiang
\thanks{M. G. Saber and Z. Jiang are with the Ottawa Research Center, Huawei Technologies Canada, 303 Terry Fox Drive, Kanata, ON, K2K 3J1, Canada.}

\thanks{Manuscript received xx, 2026; revised x xx, 2026.}}

\markboth{Photonics,~Vol.~x, No.~x, xx~2026}%
{Shell \MakeLowercase{\textit{et al.}}: Bare Demo of IEEEtran.cls for IEEE Journals}

\maketitle

\begin{abstract}
Hollow-core fiber (HCF) is often presented as an incrementally better
transmission medium to be slotted into networks designed around solid-core
silica. We argue instead that recent progress---most visibly a reported
attenuation below \mbox{0.1\,dB\,km$^{-1}$} together with a broad low-loss
window, reduced propagation delay and very low optical nonlinearity---makes it
worth asking which long-standing design conventions are intrinsic to optical
communication and which are artifacts of silica. Reviewing physical-layer,
transceiver and network-architecture implications, we suggest that the most
durable gains may come not from treating HCF as a drop-in replacement, but from
cross-layer co-design, and we outline the studies and demonstrations needed to
test where that advantage is real.
\end{abstract}


%
\IEEEpeerreviewmaketitle

\section{Introduction}

Optical fiber communication has rested on solid-core silica for half a century,
and the engineering scaffolding around it---wavelength planning, span lengths,
launch-power budgeting, digital signal processing (DSP), amplifier siting and
routing policy---co-evolved with the physical constants of that medium. Two of
those constants dominate. Silica's minimum attenuation is set by the balance between Rayleigh scattering
($\propto\lambda^{-4}$), which dominates the loss budget, and the infrared
absorption edge, placing the minimum near \mbox{1.55--1.58\,\textmu m} at
\mbox{$\sim$0.14\,dB\,km$^{-1}$} in record pure-silica-core fiber---a value
reduced only $\sim$30\% in four decades and now near silica's intrinsic limit; and the Kerr nonlinearity of glass bounds the
power a wavelength channel can usefully carry. Neither is a property of optical
communication as such---both are properties of glass. Hollow-core fiber (HCF),
in which the optical mode propagates almost entirely in air, removes the field
from the glass and therefore loosens its grip on these constants. That is a
qualitatively different proposition from refining silica, and it is the reason a
systems-level reassessment is now warranted rather than merely interesting.

Air guidance is not new~\cite{cregan1999,russell2003}, but for most of two
decades it underdelivered. Hollow-core photonic-bandgap fibers confine light
through a full two-dimensional bandgap and reached about
\mbox{1.2\,dB\,km$^{-1}$}~\cite{roberts2005}, but over narrow bandwidths, with
large polarization-mode dispersion, and with loss dominated by scattering from
the dense array of glass interfaces around the core. The decisive change was a
change of guidance mechanism. Anti-resonant fibers confine light by
anti-resonant reflection from a few thin glass membranes; nesting those
membranes---the nested anti-resonant nodeless fiber (NANF) and its double-nested
variant (DNANF)---suppresses residual leakage by orders of magnitude while
keeping the field's overlap with glass minute~\cite{poletti2014}. Because
confinement no longer requires thick or continuous glass, the bulk absorption
and Rayleigh scattering that set silica's floor are largely bypassed; the
residual loss is instead governed by confinement (leakage), by scattering from
nanometer-scale surface roughness on the membranes, and by microbending---a
different set of mechanisms with different, and in principle lower,
limits~\cite{fokoua2023}.

\begin{figure*}[t]
	\centering
	\includegraphics[width=.88\textwidth]{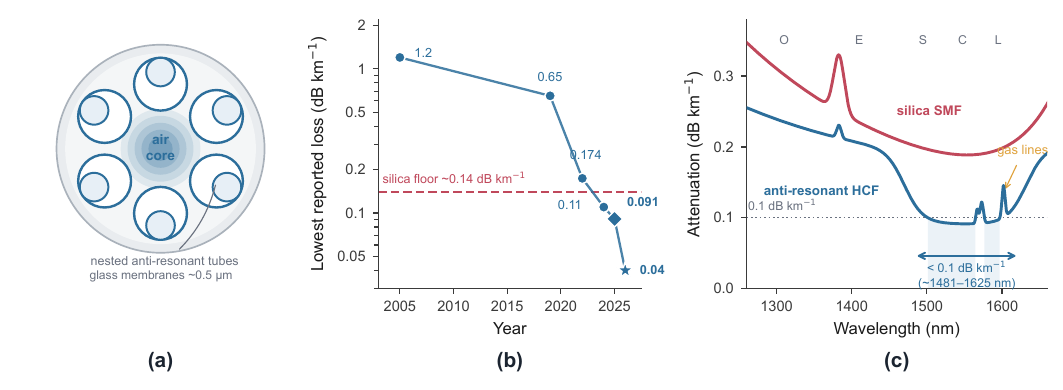}
	
	\caption{\textbf{Why hollow-core fiber has re-emerged.}
		\textbf{a},~Cross-section of a double-nested anti-resonant nodeless fiber
		(DNANF).
		\textbf{b},~Lowest reported HCF attenuation by year, with the silica record
		floor for reference (dashed)~\cite{roberts2005,bradley2019,jasion2022,chen2024,petrovich2025,li2026imi}.
		\textbf{c},~Schematic attenuation spectra of silica SMF and anti-resonant HCF
		(illustrative, not measured).}
	\label{fig:overview}
\end{figure*}

The loss trajectory follows this mechanistic shift (Fig.~\ref{fig:overview}b).
Reported attenuation fell from \mbox{0.65\,dB\,km$^{-1}$} across the C and L
bands in 2019~\cite{bradley2019} to \mbox{0.174\,dB\,km$^{-1}$} in a double-nested
design in 2022~\cite{jasion2022}, and most recently to
\mbox{0.091\,dB\,km$^{-1}$} at 1550\,nm, measured over a 15-km fiber and below
\mbox{0.1\,dB\,km$^{-1}$} across the 1481--1625\,nm window---about
18\,THz~\cite{petrovich2025}. In that single reported measurement the value sits
below silica's $\sim$0.14\,dB\,km$^{-1}$ floor, and over a markedly wider
low-loss band. The same mechanistic picture indicates where further headroom may
lie: surface-scattering loss falls approximately as the inverse cube of core
diameter and leakage falls faster still, so enlarging the core lowers both,
while microbending sets a countervailing penalty---hence an optimum core size
for each wavelength. Modeling on this basis places an optimized DNANF near
\mbox{0.04\,dB\,km$^{-1}$} and, speculatively, an ultimate limit near
\mbox{0.02\,dB\,km$^{-1}$}~\cite{fokoua2023}. These projections, and the records
themselves, warrant caution: the lowest values are reported for individual
fibers over short-to-moderate lengths, different methods (cutback versus
reflectometry) can disagree by tens of percent at these levels, and the
broad-band figures are quoted with the gas-absorption features discussed later
set aside~\cite{petrovich2025}. The substantive claim is not that any single
number is final, but that the floor which organized silica-era design no longer
self-evidently applies.

Three further properties follow from air guidance and matter as much as loss.
First, the group index is close to unity rather than $\sim$1.47, so one-way
delay is about a third lower, falling from $\sim$4.9 to
$\sim$3.3\,$\mu$s per kilometer~\cite{poletti2013}. This is a property of the
medium, not a parameter to be iterated, and is essentially unique to air
guidance. Second, because well over 99\% of the power travels in air, the
effective Kerr nonlinearity---which scales with the overlap of the optical field
with glass---is several orders of magnitude smaller than in silica; reported
nonlinear coefficients are of order \mbox{$5\times10^{-4}$\,W$^{-1}$km$^{-1}$}
against $\sim$1.3\,W$^{-1}$km$^{-1}$ for SMF, and coherent transmission has been
shown to be effectively nonlinearity-free even at watt-level
powers~\cite{liu2019,correia2026,braga2026}. Third, chromatic dispersion in the
guidance window is low and relatively flat, of order
3\,ps\,nm$^{-1}$\,km$^{-1}$---several times below standard
fiber~\cite{petrovich2025,correia2026}, although certain G.65X variants exhibit comparably low dispersion. Each of these reshapes a different part
of the design stack, as the following sections detail.

These differences arrive as latency, energy per bit and physical-layer scaling
become first-order constraints again---pushed by artificial-intelligence
training traffic, distributed data-center architectures and dense interconnect.
Industrial momentum has followed, including field and cabling trials and reports
of deployments at the scale of thousands of kilometers by hyperscale
operators~\cite{microsoft2022,ibrahimi2026,ge2026,borzycki2023}. What remains
underdeveloped is a systems-level account of which silica-era choices are
intrinsic to optical communication and which are artifacts of the medium. This
paper takes up that question. We do not claim that HCF will displace
silica or that its advantages are universal; we argue that its distinguishing
properties are now large enough, and demonstrable enough, that treating the
fiber as a faster drop-in pipe risks forfeiting most of its value.

\section*{Silica-era assumptions under reassessment}

The rules of optical network design are not arbitrary, but many encode
properties of silica rather than of communication itself. Separating the two is
the central analytical task, and it is more useful than asking whether HCF is
``better'': some conventions should survive unchanged, others invert, and the
engineering value lies in knowing which, and by how much
(Table~\ref{tab:assumptions}). Reassessment is not reversal.

Several constraints are genuinely medium-independent. Information is carried on
electromagnetic fields subject to the same quantum and thermal noise;
amplified-spontaneous-emission noise accumulates with distance and amplifier
count; coherent detection, forward error correction and the trade-off between
spectral efficiency and reach all still hold; and a channel's reach is still set
by its accumulated generalized signal-to-noise ratio (GSNR). HCF does not repeal
the Shannon limit. What it changes is the budget feeding the GSNR---how much
loss, nonlinear penalty and delay each kilometer and each amplifier contribute.

The most consequential silica-specific assumption is that the optimum launch
power is set by fiber nonlinearity. In a loaded WDM silica system, nonlinear interference acts as an additional noise source whose power grows roughly with the cube of per-channel launch power; meanwhile, the ASE contribution is approximately constant in absolute power, so its impact on SNR diminishes as signal power increases. This trade-off produces a nonlinear optimum, typically near 0–2 dBm per channel for 50~GHz channels---beyond which more power degrades performance, the familiar
``nonlinear Shannon'' picture. Cutting the nonlinear coefficient by three orders
of magnitude raises that optimum by tens of decibels, so it effectively ceases
to bind~\cite{liu2019}. Demonstrations now run boosters at 30--34.5\,dBm and
convert the resulting GSNR headroom into reach or spectral
efficiency~\cite{mardoyan2026,boddeda2026}. A subtler consequence is the
decoupling of per-channel power. In silica the nonlinear interference is
intertwined across the comb---raising one channel's launch power adds
nonlinear penalty to its neighbors, so a channel
cannot be tuned without degrading the rest. In HCF the negligible nonlinearity
removes this coupling, so each channel's power becomes a near-independent control
variable, which is precisely what makes per-channel transmitter-power
optimization an effective way to equalize performance across a fully loaded
link~\cite{hong2026b}. The constraint does not vanish; instead, it
migrates to amplifier output power and efficiency, to optical damage and
connectorization at the solid-core interfaces that bound an HCF span, and to the
gas-absorption and intermodal interference effects, and to the transceiver noise. Tellingly, network analyses already
find that the best in-line amplifier output power is not the maximum available.
From an electrical power consumption perspective, the optimum sits near 26~dBm, with diminishing or negative returns beyond roughly 32~dBm at a loss of 0.11~dBkm$^{-1}$~\cite{zami2026}. From a nonlinearity standpoint, the optimum shifts to around 37~dBm when considering the markedly reduced nonlinear coefficient of HCF, $\gamma \approx 5\times 10^{-4}\,\mathrm{W^{-1}\,km^{-1}}$ at a loss of 0.1~dBkm$^{-1}$~\cite{9669036}.

The second assumption is spectral. Silica's low-loss window near 1.55\,$\mu$m is
the crossover of Rayleigh scattering (falling as $\lambda^{-4}$) and infrared
absorption (rising with wavelength)---a property of the glass, and the reason
the amplifier and component ecosystem clusters in the C and L bands. In
anti-resonant HCF the window is set instead by membrane thickness through the
anti-resonance condition, decoupling it from material absorption; low loss has
been measured across $\sim$18\,THz and can in principle be placed from
$\sim$1\,$\mu$m to beyond 2\,$\mu$m, with dual-band designs already
demonstrated~\cite{sakr2020,mahdiraji2026,petrovich2025}. Whether this freedom is
usable depends entirely on whether amplifiers and components exist for the
chosen band---which today, outside the silica window, they largely do not. The
third assumption, that propagation delay is fixed by the medium, gives way once
$n_g\!\approx\!1$: delay becomes partly an engineering choice and, in a
heterogeneous network, a per-path one. The fourth, that DSP must devote
substantial resources to chromatic-dispersion compensation, weakens because the
equalizer length scales with accumulated dispersion, which is several times
lower and flatter here; the net receiver-DSP saving is real but bounded by the
carrier-recovery, adaptive-equalization and FEC blocks that dominate
complexity~\cite{petrovich2025,saber2026practicallimit,correia2026}.

Finally, some assumptions are incomplete rather than wrong. Silica SMF is
treated as a near-ideal, weakly scattering, single-mode, low-reflection
waveguide. HCF is better described as effectively single-mode: higher-order
modes exist but are differentially attenuated, and their residual coupling
produces intermodal interference (IMI)---a coherent, multipath impairment with
differential group delays of order nanoseconds per kilometer and no close analog
in mature SMF links~\cite{fokoua2023,fontaine2026}. Backscatter is roughly 40\,dB
below that of SMF, which suppresses the coherent crosstalk that otherwise limits
single-fiber bidirectional transmission, but simultaneously weakens the Rayleigh
signal on which reflectometry depends~\cite{nakamura2026}. Although bidirectional transmission has been demonstrated in laboratory settings, the larger reflections at SMF--HCF interfaces may render multipath interference (MPI) a practical limitation in field deployments, an issue that remains insufficiently studied. And because the core is hollow, trace CO$_2$ and
water vapor imprint narrow absorption lines on an otherwise flat spectrum---an
impairment with no counterpart in solid glass~\cite{li2026co2,sillekens2026}. Air
guidance thus redistributes both advantages and impairments; a framework
inherited wholesale from silica will neither exploit the former nor anticipate
the latter.

\begin{figure*}[t]
	\centering
	\includegraphics[width=.99\textwidth]{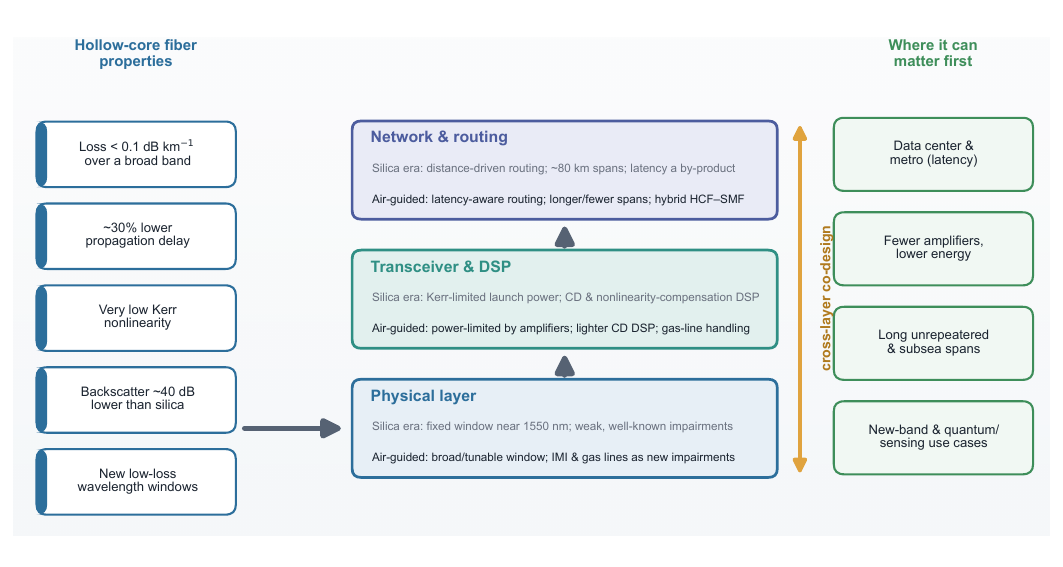}
	\caption{\textbf{From a faster fiber to cross-layer co-design.} Conceptual
		schematic: hollow-core fiber properties (left) propagate through the physical,
		transceiver/DSP and network layers (center)---each with a silica-era assumption
		(gray) and an air-guided reconsideration (black)---to where HCF can help first
		(right). The vertical arrow denotes cross-layer coupling.}
	\label{fig:concept}
\end{figure*}

These shifts also refuse to stay within one layer, which is why piecemeal
adoption tends to disappoint. A fiber-level change---lower nonlinearity, lower
delay, a new gas-line impairment---yields value only if the transceiver, DSP and
network layers are adjusted to exploit or absorb it: high launch power is wasted
without amplifiers and interfaces that tolerate it; a broad low-loss band is
inert without matching amplification; lower latency is invisible to a routing
layer that does not treat delay as a variable. Figure~\ref{fig:concept} makes
this coupling explicit, tracing how the distinguishing properties of HCF
propagate from physical-layer design through transceiver and DSP choices to
network and routing policy. This cross-layer view---not any single record
metric---is the throughline of the two sections that follow.

\begin{table*}[t]
	\centering
	\caption{\textbf{Selected silica-era design assumptions and their status in an
			air-guided regime.} The intent is to make dependencies explicit, not to claim
		that every convention should change. Entries are qualitative and indicate
		direction rather than settled conclusions.}
	\label{tab:assumptions}
	\small
	\begin{tabularx}{\textwidth}{p{4.6cm} Y Y}
		\toprule
		\textbf{Design assumption} & \textbf{Basis in silica systems} &
		\textbf{Status for anti-resonant HCF} \\
		\midrule
		Optimum launch power is set by Kerr nonlinearity &
		Nonlinear interference grows with power, creating a clear optimum &
		\emph{Reassess:} nonlinearity is orders of magnitude lower; limits shift to
		amplifier power, interfaces and HCF-specific effects~\cite{liu2019} \\
		Usable spectrum is the silica window near 1550\,nm &
		Loss minimum and mature amplifiers sit in the C/L bands &
		\emph{Reassess:} low-loss guidance is broader and tunable, but amplifiers and
		components for new windows are immature~\cite{sakr2020,mahdiraji2026} \\
		Propagation delay is fixed by the medium &
		Group index $\sim$1.47 is a material constant &
		\emph{Reassess:} delay is $\sim$30\% lower and becomes partly a design
		choice~\cite{poletti2013} \\
		DSP must heavily manage chromatic dispersion &
		Standard fiber has substantial accumulated dispersion &
		\emph{Reassess:} dispersion is low and flatter in the guidance
		window~\cite{petrovich2025,correia2026} \\
		Fiber is a near-ideal single-mode, low-reflection waveguide &
		SMF is effectively single-mode with weak, well-characterized scatter &
		\emph{Reassess:} effectively (not strictly) single-mode; IMI and very low
		backscatter change monitoring and transmission~\cite{fokoua2023,nakamura2026} \\
		Reach is limited by accumulated SNR; Shannon limit applies &
		Fundamental noise and capacity limits &
		\emph{Retain:} unchanged by air guidance \\
		\bottomrule
	\end{tabularx}
\end{table*}

\section*{Physical-layer and transceiver implications}

\begin{figure*}[t]
	\centering
	\includegraphics[width=.95\textwidth]{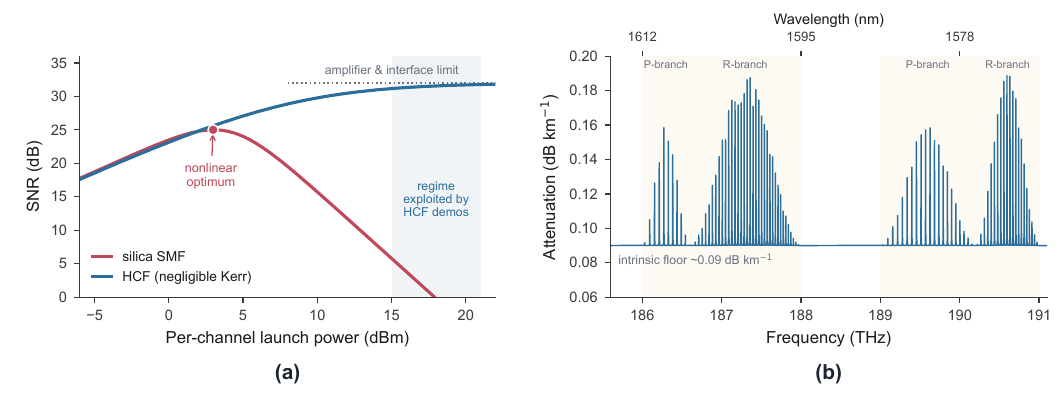}
	\caption{\textbf{Physical-layer trade-offs under air guidance}.
		\textbf{a},~SNR versus per-channel launch power for silica SMF and
		HCF~\cite{liu2019} (illustrative, not measured).
		\textbf{b},~Modeled CO$_2$ absorption comb in the L band, showing the P- and
		R-branches of two CO$_2$ rotational-vibrational bands; loss per unit length at a
		0.10\,dB\,km$^{-1}$ R-branch peak and 1\,GHz
		linewidth~\cite{li2026co2,he2026,wang2025co2} (illustrative, not measured).}
	\label{fig:tradeoffs}
\end{figure*}

\textbf{Launch power, modulation and DSP.}
The lifted nonlinear ceiling reshapes the joint choice of launch power, modulation
format and reach (Fig.~\ref{fig:tradeoffs}a). With nonlinear interference suppressed,
the SNR-versus-power curve loses the peak that defines silica operation over the
accessible power range, and the additional headroom has been spent in two ways.
One is spectral efficiency: probabilistically shaped 64- and 256-QAM, supported
by ring-wise neural-network equalization, has carried an aggregate
$\sim$0.55\,Pb\,s$^{-1}$ across the S, C and L bands on a single
fiber~\cite{li2026}. The other is reach at high power: boosted hybrid and
unrepeatered spans of hundreds of kilometers~\cite{mardoyan2026,boddeda2026}. The
lesson is not that more power is always better. The optimum migrates to amplifier
saturation, efficiency and interface damage, and at network scale the
economically optimal in-line power can sit well below the maximum---near 26\,dBm
in one transparent-network study, with no benefit, and eventually harm, beyond
$\sim$32\,dBm~\cite{zami2026, saber2026_hcfimdd}. On DSP, the chromatic-dispersion equalizer is the
one block whose length scales directly with accumulated dispersion; at
$\sim$3\,ps\,nm$^{-1}$km$^{-1}$ it shrinks by roughly the dispersion ratio
relative to standard fiber. But carrier recovery, adaptive equalization and FEC
are unchanged, and---as the next paragraphs show---IMI and gas lines can
add equalizer taps, so the net receiver-DSP complexity of an HCF link is
not obviously lower and should be measured end-to-end rather than assumed.
A further limit sits in the transceiver itself. Air guidance can supply ample
link GSNR, but the back-to-back signal-to-noise ratio of the transponder---set
by digital-to-analog and analog-to-digital converter resolution, driver and
modulator noise, and component bandwidth---degrades as the symbol rate rises,
because the same impairments are spread over a wider band and the effective
number of bits falls at high frequencies. At high baud the
transceiver noise floor, not the fiber, can cap the achievable SNR, so the
headroom HCF opens at the link level need not translate into higher-order
formats at arbitrarily increased baud. The practical operating point is therefore a
joint optimization of baud rate, modulation order and reach against the
transceiver noise floor, rather than a link-SNR calculation
alone~\cite{klaus2022,saber2026practicallimit}.

\textbf{Bend sensitivity, mode and polarization control.}
Several properties are reshaped rather than removed, and they trade against one
another. Lower loss favors a larger air core, because both leakage and
surface-scattering loss fall with core size; but microbending loss rises with
core size, and so does bend sensitivity. The same core enlargement that modeling
associates with pushing loss toward
\mbox{0.03--0.02\,dB\,km$^{-1}$} also raises the radius at which bending becomes
lossy, so that loss, bandwidth and bend tolerance cannot be maximized
independently---they define a multi-objective optimum that depends on the
deployment (tight data-center routing versus long, gently bent
spans)~\cite{fokoua2023,petrovich2025}. Mode control is likewise a design target,
not a given: guidance is effectively, not strictly, single-mode, so a fiber must
be engineered to differentially attenuate higher-order modes (often quantified by
a higher-order-mode extinction ratio), and this competes with loss and bandwidth
objectives, while bend- and splice-induced coupling can still dominate the IMI a
deployed link actually sees~\cite{li2026imi,fontaine2026}. Polarization behavior
also differs: recent anti-resonant fibers report low polarization-mode
dispersion ($\sim$0.1\,ps\,km$^{-1/2}$), unlike early bandgap fibers, but the
phase and polarization dynamics of deployed cable---relevant to coherent
receivers and to sensing---are only beginning to be characterized in the
field~\cite{petrovich2025,correia2026,fang2026}.

\textbf{Intermodal interference: the impairment that replaces nonlinearity.}
Reduced nonlinearity does not leave a clean channel; it exposes a different limit.
Residual coupling to higher-order modes produces IMI, a coherent multipath effect
that appears in the channel's impulse response as a delayed plateau trailing the
fundamental mode, with differential group delays of order
4--5\,ns\,km$^{-1}$~\cite{fontaine2026}. Reported IMI in recent low-loss fibers
spans roughly $-50$ to below $-70$\,dB\,km$^{-1}$, and modeling indicates that
levels below about $-60$\,dB\,km$^{-1}$ are needed for transoceanic
reach---a threshold met in a fiber measured at $-68.8$\,dB\,km$^{-1}$ that enabled
6660\,km transmission~\cite{li2026imi,boddeda2026,fokoua2023}. Two features make
IMI strategically different from Kerr noise. First, it does not improve by
lowering power; it is fixed by fiber and splice quality. Second, it accumulates
with the worst segment, so a single non-uniform span or poor splice can set
the IMI of an entire link, placing a premium on manufacturing uniformity and
splice control~\cite{li2026imi}. Mitigations therefore act on the fiber (stronger
higher-order-mode suppression) and on the DSP: digital subcarrier multiplexing
narrows each subcarrier so it has been claimed to be resilient to the frequency-selective fading IMI produces, extending reach at comparable complexity~\cite{ospina2026}.

\textbf{Gas-line absorption: a new, wavelength-selective limit.}
Because the core is hollow, residual CO$_2$ and water vapor imprint their
rotational-vibrational lines on the transmission spectrum as narrow notches---of order a
gigahertz wide and, in the strongest lines, up to $\sim$0.5\,dB\,km$^{-1}$
deep---on an otherwise flat
background (Fig.~\ref{fig:tradeoffs}b)~\cite{li2026co2,he2026,wang2025co2}. This is qualitatively
unlike any silica impairment: the loss is channel-selective rather than
band-wide, and because it scales with length a notch that is negligible over a
few kilometers can exceed 10\,dB over hundreds, removing specific channels rather
than degrading all of them uniformly. Mitigations form a layered toolkit with
quantified trade-offs. Fiber- and process-level methods reduce gas content or
seal the core under positive pressure. At the transmitter, spectral
pre-emphasis has extended reach from 150 to $\sim$300\,km at a 1\,dB penalty in
one study, and frequency-domain spectral pre-equalization has recovered a 10\,dB
notch using three equalizer taps---about 5.5\,dB better than a 383-tap adaptive
equalizer~\cite{li2026co2,sillekens2026}. Per-channel transmitter-power
optimization has equalized performance across a fully loaded
$32\times800$\,Gb\,s$^{-1}$ C-band link over 442\,km, holding every channel above
a 1.5\,dB Q-margin~\cite{hong2026b}; adaptive baud-rate and subcarrier allocation
that step signal energy around the lines have been used out to
6660\,km~\cite{boddeda2026}; and efficient in-field measurement can estimate the
resulting penalty for deployment planning~\cite{he2026}. None removes the effect
entirely---each trades against usable bandwidth or transmitter complexity---and
the right combination is length- and band-specific~\cite{photonics13060559}.

\textbf{Wavelength windows and the component ecosystem.}
If the low-loss window is no longer tied to silica, new operating bands become
conceivable: dual-band anti-resonant fibers with low loss at both $\sim$1\,$\mu$m
and $\sim$1.55\,$\mu$m have been demonstrated, and fully integrated 1064\,nm
transmitters built explicitly to exploit such windows~\cite{mahdiraji2026,shen2026}.
Realizing them depends on a component ecosystem that does not yet match C-band
maturity---most pressingly, amplification for new or wider bands, and low-loss,
low-reflection interfaces. The interface problem is concrete: the lowest-loss
HCFs have mode-field diameters roughly twice that of SMF, and the air--glass
boundary presents a $\sim$3.5\% Fresnel reflection (about 0.155\,dB) that, left
untreated, both wastes power and seeds multipath echoes~\cite{slavik2026,
	zandueta2026}. Progress is real but recent: homogeneous HCF--HCF splices average
$\sim$0.05\,dB with field-deployable automated alignment in under 100\,s at full
yield across repeated trials, and HCF--SMF coupling below 0.2\,dB with
back-reflection under $-60$\,dB has been reached with mode-field adapters or
lensed interfaces~\cite{zhang2026splice,feng2026,slavik2026,zandueta2026}. On the
manufacturing side, in-line interferometric measurement with closed-loop pressure
control has held capillary dimensions to about $\pm$1\% over a 20-km draw, a
prerequisite for the uniformity that IMI demands~\cite{budd2026}. These
interface and yield economics, more than fiber loss, gate cost and reliability,
and are likely to determine where HCF is adopted first.

\textbf{Monitoring and sensing.}
The $\sim$40\,dB lower backscatter is doubly consequential~\cite{nakamura2026}. It
suppresses the coherent crosstalk that limits same-fiber bidirectional
transmission, which several demonstrations
exploit~\cite{hong2026,dupas2026}; but it blinds the optical
time-domain reflectometry that operators rely
on for fault location and live diagnostics. Because the HCF itself has negligible
nonlinearity, longitudinal power-profile estimation must instead lean on the
short solid-core jumpers embedded at amplifier sites, which still provide a
usable nonlinear signature~\cite{hahn2026,fontaine2026}. Pilot-tone-based
monitoring, by contrast, becomes cleaner. Here a low-frequency, small-amplitude
tone is superimposed on each data channel; in silica its accuracy is eroded by
stimulated-Raman-scattering tone transfer between channels and by
chromatic-dispersion-induced tone fading---two effects that push the optimal tone
frequency in opposite directions and force a compromise. Air guidance removes the
former through its negligible nonlinearity and suppresses the latter through its
low dispersion, so the tone can be recovered without that
trade-off~\cite{jiang2022pt}. Sensing inverts the usual hierarchy: the same low backscatter and roughly order-of-magnitude lower thermal sensitivity make HCF a poor distributed temperature sensor, although distributed sensing in unmodified NANF and in deployed field cables has already been demonstrated~\cite{chen2026sensing,fang2026}. The general point is that HCF
does not simply improve on silica's auxiliary functions---it changes them, and
the monitoring and sensing stack must be redesigned accordingly.

\section*{Network architecture and routing}

\begin{figure*}[t]
	\centering
	\includegraphics[width=\textwidth]{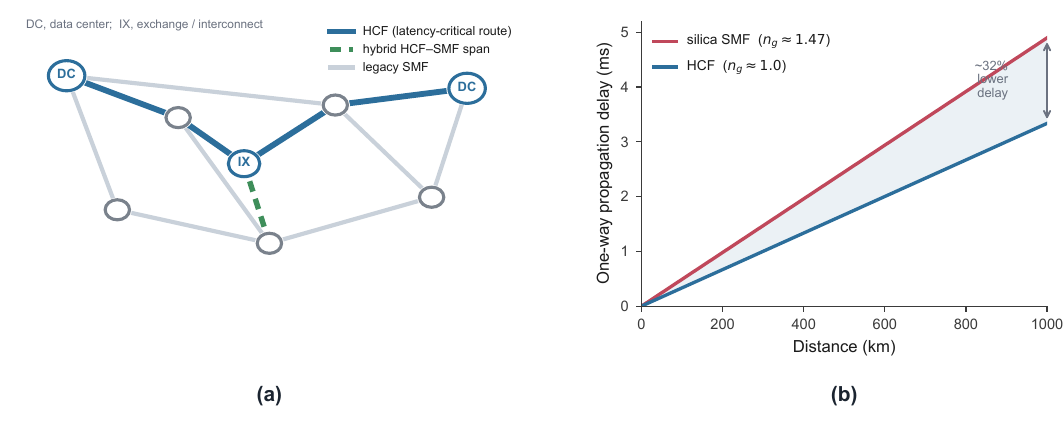}
	\caption{\textbf{From links to networks.}
		\textbf{a},~Selective HCF deployment on a heterogeneous network: latency-critical
		routes and hybrid HCF--SMF spans alongside legacy SMF.
		\textbf{b},~One-way propagation delay versus distance for silica SMF and
		HCF~\cite{poletti2013}.}
	\label{fig:network}
\end{figure*}

The properties that reshape the physical layer also bear on where HCF is
deployed and how networks are planned. Three regimes stand out as early
candidates for decisive advantage, and one cross-cutting idea---treating latency
as a design variable---runs through them (Fig.~\ref{fig:network}).

\textbf{Where HCF may help first.}
The clearest near-term case is the latency-bounded, distance-limited link.
Data-center interconnect and metro routes turn roughly 30\% lower propagation
delay directly into application value---tighter clock synchronization,
lower round-trip times for distributed storage and consensus, and higher
goodput for the latency-gated collective operations that increasingly dominate
artificial-intelligence training---while their short reach keeps the amplifier
and interface count modest~\cite{ibrahimi2026,poletti2013}. Bidirectional,
full-band transmission over field-deployable HCF cable using commercial coherent
pluggables has been shown, making this regime concrete rather than
hypothetical~\cite{hong2026,dupas2026}. A second case is the
amplifier- and energy-constrained route. Below \mbox{0.1\,dB\,km$^{-1}$}, a span
of fixed length incurs roughly half the loss of an SMF span, or equivalently the
span can be doubled in length for the same loss, halving the number of in-line
amplifier sites; transparent-network analyses translate this into amplifier
power savings of order tens of percent per Tb\,s$^{-1}$ and into capacity gains,
with one study reporting up to a doubling of maximum network capacity and a
$\sim$35\% reduction in energy per transported bit when long, low-loss spans are
combined with an amplifier power optimum near 26\,dBm~\cite{ibrahimi2026,zami2026}.
A third, more forward-looking case is the long unrepeatered or sparsely
repeatered span. Hybrid HCF--SMF spans exceeding 200\,km have delivered
$>$800\,Gb\,s$^{-1}$ achievable information rate over 1113\,km, and a
sparsely repeatered transoceanic experiment reached 6660\,km using 266-km spans
with fewer than 30 repeaters---against the $\sim$100 of a conventional system---
enabled jointly by low IMI ($-68.8$\,dB\,km$^{-1}$), high-power boosting and
adaptive baud-rate management around the gas lines~\cite{mardoyan2026,boddeda2026}.
These remain laboratory or trial results, frequently in recirculating loops, and
their translation to fielded systems is not yet established.

The breadth of demonstrations is itself the signal. Across the network
they now span access (bidirectional coherent passive optical networking over
anti-resonant HCF at 200/50\,Gb\,s$^{-1}$ down/up), the data-center and metro
tier (wavelength-reconfigurable optical switching at multi-petabit aggregate
scale), and the high-capacity core ($\sim$0.55\,Pb\,s$^{-1}$ on one fiber, and
real-time 2\,Tb\,s$^{-1}$-class transponders over a 120-km HCF span with latency
comparable to 80\,km of SMF)~\cite{liu2026pon,huang2026,li2026}. Taken
together---and with the caveat that most are trials---these indicate that raw
capacity is not the binding constraint. The open questions concern cost,
interfaces, impairments and day-to-day operation, which shifts the center of
gravity of the problem from the fiber to the system and the network.

\textbf{Hybrid silica--HCF networks.}
For the foreseeable future networks will be heterogeneous~\cite{saber2026ecoc,saber2026protectionswitchinghybridhollowcore}, with HCF introduced
selectively alongside an installed silica base, so the design question is which
links or spans to convert under a fixed budget. The quantitative answers are
encouraging but assumption-dependent. GSNR-based studies on realistic carrier
topologies report HCF advantages of up to $\sim$8\,dB and carried-traffic gains
of order ten percent over an all-SMF baseline, bracketed by the assumed loss,
splice quality and amplifier configuration~\cite{correia2026}. Time-domain
modeling of multi-span systems finds that 25--50\% of the HCF can be removed in
favor of SMF with little performance loss in metro and unrepeatered regimes,
because much of HCF's benefit comes from loss and high-power tolerance that a
hybrid span can capture with less fiber~\cite{braga2026,mardoyan2026}. And
integer-programming placement studies show feasible-path counts for 800\,Gb\,s$^{-1}$
rising by about 36\% when only 10\% of spans are HCF and by up to 100\% at a 20\%
budget, with 85\% of feasible paths reachable when 25--55\% of spans are
converted~\cite{pedro2026}. The consistent message---that a minority of
well-chosen HCF can capture most of the benefit---reframes deployment as a
budgeted optimization rather than wholesale replacement, but the specific
fractions are sensitive to the input assumptions and should guide experiments
rather than substitute for them.

\textbf{Latency as a first-class routing variable.}
Conventional routing and wavelength assignment optimize for cost, capacity and
reach, treating propagation latency as a by-product of the chosen path. That is
reasonable in an all-silica network, where every fiber kilometer carries the same
group index and therefore the same delay. Heterogeneity breaks the assumption:
two paths of equal physical length can differ in latency according to how much of
each traverses air-guided fiber (Fig.~\ref{fig:network}b), so delay becomes a
controllable, per-path quantity. This invites treating latency as an explicit
routing metric or constraint alongside capacity and cost---most valuably for
traffic with hard delay bounds: the synchronization and collective-communication
phases of distributed AI training, where tail latency can gate cluster
throughput; financial messaging; industrial control; and distributed consensus~\cite{saberAIdata}.
Delay-constrained routing is itself well studied, but it has not had to contend
with a physical layer in which the delay per kilometer varies by medium,
nor with the coupled decision of where to place a limited budget of
low-latency fiber. Network studies that begin from latency or energy budgets
already hint at the payoff: selectively upgrading under a fifth of links can cut
edge-data-center consolidation cost by tens of percent under latency bounds, and
similar selective placement reduces the resources needed for quantum-key
distribution~\cite{ibrahimi2026,pedro2026}. We advance this as a hypothesis to be
tested, not a settled result: the open question is whether planning tools and
control-plane abstractions that treat fiber type, latency and energy as joint
variables can realize, on real traffic, the advantage the physical layer now
makes available.

\section*{Roadmap and open questions}

If HCF is to be more than a faster pipe, the community needs evidence on where
air guidance changes the right engineering answer, not merely where it improves a
fiber metric. We see five concrete priorities.

\emph{Standardized, independent characterization.} The headline loss, IMI and
gas-absorption figures are reported for individual fibers and lengths, sometimes
with key quantities (the lowest losses, the wide-band projections) derived from
modeling rather than measured~\cite{petrovich2025,fokoua2023}. The field needs
agreed methods---loss measured over tens of kilometers rather than extrapolated,
IMI and differential mode attenuation, backscatter, and gas-line spectra under
controlled fill---together with independent cross-checks and round-robin
comparisons. Standardization activity in this direction has begun and deserves
support~\cite{ge2026}; without it, system designers cannot use the numbers with
confidence.

\emph{System studies that price the new impairments.} The benefits of low
nonlinearity, low delay and low dispersion must be weighed against IMI, gas-line
absorption and interface loss in fully loaded, multi-span links representative of
deployment---not single-channel or recirculating-loop experiments, which can
flatter both the benefits and the impairments. The most useful studies will
report the net receiver-DSP and amplifier complexity of an HCF system, so
that the equalizer taps saved on dispersion and the taps spent on IMI and gas
lines are counted together, and so that high-power operation is costed against
amplifier efficiency and reliability~\cite{ospina2026,hahn2026,zami2026}. A
parallel question is whether the modulation format itself can be chosen to
minimize the new impairments rather than merely tolerate them: digital subcarrier
multiplexing, OFDM and other multicarrier or adaptively shaped waveforms can
narrow and steer subcarriers around gas lines and spread energy to resist the
frequency-selective fading IMI produces. A systematic comparison of these
formats---on the joint axes of IMI and gas-line resilience, reach and DSP
cost---would help identify which waveform is best matched to an air-guided
channel~\cite{ospina2026,wang2025co2}.

\emph{The interface and component ecosystem.} Field-grade, low-loss,
low-reflection HCF--SMF interfaces, fast and reliable splicing, manufacturing
uniformity sufficient to keep link-level IMI below threshold, and amplification
matched to wider or new bands are prerequisites for most use cases---and, as
argued above, their economics may set adoption timing more than fiber loss
does~\cite{slavik2026,zhang2026splice,budd2026,mahdiraji2026}. New-band
amplification, in particular, is the gating technology for exploiting HCF's
broader low-loss window at all; candidate gain media exist---ytterbium near
1\,$\mu$m, bismuth in the O, E and S bands, thulium and holmium near
2\,$\mu$m---but none is mature at telecom scale~\cite{petrovich2025}. Two
further unknowns belong on the same list: absolute cable cost, for which
little public data exist while manufacturing scales up, and long-term
reliability---aging of the thin membranes, gas or moisture ingress at cable
breaks, and field repair---which remains essentially uncharacterized.

\emph{Cross-layer co-design and planning.} The hybrid-deployment and
latency-as-metric questions call for planning tools and control-plane
abstractions that treat fiber type, latency and energy as joint optimization
variables rather than post-hoc attributes. The decisive evidence would be
field-representative testbeds that route real delay-bounded traffic over mixed
HCF--SMF paths and demonstrate, end to end, that latency-aware placement and
routing deliver the predicted gains~\cite{braga2026,pedro2026,ibrahimi2026}.
Several directions are actionable now. First, extend GSNR-based
routing-and-wavelength-assignment engines so that fiber type, group delay and
energy per bit are carried as first-class link weights, rather than attributes
recovered after the path is fixed; on hybrid routes this means making the engine
explicitly fiber-transition-aware, modeling each SMF--HCF junction as its
own element so that the interface loss and back-reflection it adds, the
segment-by-segment change in dispersion, and the IMI and gas-line penalties that
accrue only on the air-guided portions are all priced into the path metric---and
so that a route is penalized for the number of transitions it incurs, not
merely for the fraction of HCF it traverses. Second, frame selective deployment as a
budgeted placement problem---ranking candidate spans by marginal benefit per unit
cost (latency removed, amplifier sites saved, feasible high-rate paths gained)
and converting greedily under explicit capital-expenditure and energy
ceilings~\cite{pedro2026}. Third, agree on a control-plane data model that exposes
fiber type, accumulated IMI and the per-channel gas-line map to the planner, so
that the routing and optical layers act on a shared view of the link. Finally,
because launch power, modulation format and route are coupled once the medium
varies along a path, these should be optimized jointly rather than in sequence---and
that joint optimization must treat the transceiver back-to-back SNR as a hard
ceiling, since the GSNR headroom HCF opens at the link level is wasted on any route
that is already transceiver-limited. The co-design response is to spend that headroom
where the converters can use it: rather than chasing record symbol rates the
digital-to-analog and analog-to-digital converters cannot support, exploit HCF's wide
low-loss window for parallelism---many moderate-baud subcarriers across the band
rather than a few ultra-high-baud carriers---combined with probabilistic
constellation shaping and digital pre-distortion, so that the planner directs GSNR
headroom to the routes and formats where it actually converts to delivered capacity.

\emph{Beyond transmission.} Higher delivered power, lower loss and new
wavelength windows also bear on adjacent functions---power-over-fiber beyond the
fiber-fuse limit of silica, wider repeater spacing and memory-native wavelengths
for quantum networking, and distributed sensing with behavior distinct from
silica---where the early evidence is promising but largely model- or
proof-of-concept-based~\cite{mcculloch2026,mantri2026,chen2026sensing}. Because
these uses may favor different fiber designs and wavelengths, they could
influence which variants the ecosystem ultimately standardizes on.

The honest summary is that HCF stands at an inflection point but is not yet
mature. Its distinguishing properties are real and, for the first time,
demonstrable at system scale; its impairments are equally real and less
familiar; and most of the decisive evidence remains to be gathered in
fully loaded, fielded systems. The constructive stance is neither uncritical
enthusiasm nor dismissal as a niche medium, but a deliberate program of
cross-layer co-design and independent measurement to locate, quantitatively,
where air guidance changes the answer---and where it does not. Treated as a
catalyst for re-examining inherited assumptions rather than as a drop-in
replacement, HCF is most likely to yield an advantage that endures.

\section*{Disclaimer and Acknowledgment}
The views and opinions expressed in this article are solely those of the author and do not reflect the official position of Huawei. Generative AI was used for language and grammatical refinement.

\ifCLASSOPTIONcaptionsoff
  \newpage
\fi

\bibliographystyle{IEEEtran}
\bibliography{references}

\end{document}